\def\selectedoptions{final}

\documentclass[\selectedoptions]{aipproc}

\layoutstyle{8x11double}

\newcommand{\beq}{\begin{equation}}
\newcommand{\eeq}{\end{equation}}

\begin{document}

\title 
      [IMPLICATIONS OF SHOCK WAVE EXPERIMENTS WITH PRECOMPRESSED MATERIALS FOR GIANT PLANET INTERIORS]
      {
       {\small \sf \vspace*{-65pt}Submitted as proceedings article for the American Physical 
        Society meeting on\\[-5pt] Shock Compression of Condensed Matter, Hawaii, June, 2007.\\[40pt]
       }
       IMPLICATIONS OF SHOCK WAVE EXPERIMENTS\\ WITH PRECOMPRESSED MATERIALS FOR\\ GIANT PLANET INTERIORS 
      }


\author{Burkhard Militzer}{
  address={Geophysical Laboratory, Carnegie Institution of Washington,
5251 Broad Branch Road, NW, \\ Washington, DC 20015},
  email={militzer@gl.ciw.edu},
  thanks={}
}

\iftrue
\author{William B. Hubbard}{
  address={Lunar and Planetary Laboratory, University of Arizona, Tucson, AZ 85721},
  email={hubbard@lpl.arizona.edu},
}
\fi

\copyrightyear  {2007}

\begin{abstract}
This work uses density functional molecular dynamics simulations of
fluid helium at high pressure to examine how shock wave experiments
with precompressed samples can help characterizing the interior of
giant planets. In particular, we analyze how large of a precompression
is needed to probe a certain depth in a planet's gas envelope. We
find that precompressions of up to 0.1, 1.0, 10, or 100 GPa are needed
to characterized 2.5, 5.9, 18, to 63\% of Jupiter's envelope by mass.
\end{abstract}

\date{\today}

\maketitle

\section{INTRODUCTION}

Shock wave experiments have served a the primary experimental
technique to study material at high pressure and temperature. Laser
shocks~\cite{Si97,Co98} as well as magnetically driven
shocks~\cite{Kn01,Kn03} enable us to reaches megabar pressures and
temperature of tens of thousands of degrees Kelvin. The main advantage
of this technique is that it requires one to measure only the velocity
of the shock front, $u_s$, and that of the impactor, $u_p$, in order
to obtain direct information about the equation of state. The
conservation of mass, momentum, and energy~\cite{Ze66} across the
shock front relates the initial ($E_1,P_1,V_1$) and the final
($E_2,P_2,V_2$) internal energies, pressures and volumes of the
material,
\begin{eqnarray}
P_2 - P_1 &=& \rho_1 u_s u_p~~~,~~~\\
\frac{\rho_2}{\rho_1} &=& \frac{u_s}{u_s-u_p}~~~,~~~\\ 
(E_2-E_1) \!\!\!\!&+& \!\!\!\! \frac{1}{2} (P_2+P_1)(V_2-V_1) = 0~~~.
\label{hug}
\end{eqnarray}
It is assumed that the shocked material reaches thermal equilibrium
during the experiment that typically last on the order of
nanoseconds. This assumption is well justified in most cases unless
the shock triggers a phase transformation, e.g. freezing under shock
loading, or a slow chemical reaction that is a bit more complex that
the mere dissociation of molecules, which occurs very fast.

\begin{figure}
\includegraphics[width=7.0cm]{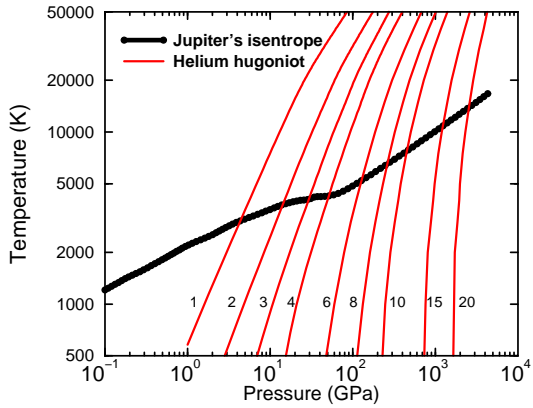}
\caption{Single shock helium Hugoniot curves for 
different precompression ratios, $V_0/V_1$ as indicated by the labels,
are compared with Jupiter's isentrope.}
\label{jupiter} 
\end{figure}

One disadvantage of shock experiments is that one cannot control the
temperature independently from the shock pressure. The velocity of the
impactor, also called particle velocity, and the equation of state of
the sample material uniquely determine the final temperature,
pressure, and density. The collection of all final states that can be
reached for different particle velocities is called a Hugoniot curve
(Fig.~\ref{jupiter}). 

Initially the compression ratio increases with the particle velocity
but above a certain value, most of the shock energy is converted to
heating the sample rather that compressing it. For the main
constituents in giant gas planets, hydrogen and helium, the maximum
compression ratios, $V_1/V_2$, are 4.3~\cite{MC00} and
5.24~\cite{Mi06} respectively. The compression ratio rarely reaches
values that are much than that. The maximum shock compression ratio is
controlled by the balance of excitations of internal degrees of
freedom, which increase the compression, and interaction effects that
reduce the compression~\cite{Mi06}.

As a result of this limitation for the attainable density, it becomes
rather difficult to characterize the interior of planets with shock
experiments alone. The shock Hugoniot curves rise much faster in a
temperature-pressure diagram than planetary
isentropes. Figure~\ref{jupiter} illustrates this for helium as a
sample material. 

This impasse can now be addressed with a new experimental
technique~\cite{Lee06} that combines static and dynamic
compression. By first compressing the sample statically in a diamond
anvil cell, the starting density can be sufficiently increased so that
a subsequent shock experiment yields equation of state data along a
different Hugoniot curve at higher density. Although the compression
ratio has been shown to be reduced asa result of the
precompression~\cite{Mi06}, the absolute densities are of course
higher due to the precompression. Therefore a larger section of the
giant planet interiors can be studied. The purpose of the article is
to understand quantitatively how much of a precompression is needed to
characterize a substantial part of Jupiter's gaseous envelope.

\section{METHODS}
To characterize the properties of helium at high pressure and
temperature, we use density functional molecular dynamics (DFT-MD)
computer simulation that we perform with the Vienna Ab-initio
Simulation Package~\cite{VASP}. The simulation were performed with 64
atoms in the unit cell using Born Oppenheimer molecular dynamics that
derive the instantaneous forces from an electronic structure
calculation. We used the Perdew-Burke-Ernzerhof generalized gradient
approximation~\cite{PBE} for the exchange-correlation energy, and
$\Gamma$-point sampling of the Brillouin zone. Since electronic
excitation are important to characterize helium Hugoniot curve above
10$\,$000$\,$K~\cite{Mi06}, we used a finite temperature Mermin
functional to model electronic excitations in thermal
equilibrium. More details about the simulation and discussion of
finite size corrections can be found in
Refs.~\cite{Mi06,Vo07b,Vo07}. We derived the equation of state of
dense fluid helium by performing DFT-MD simulations for large grid of
density-temperature points and obtained the shock Hugoniot curves by
solving Eq.~\ref{hug}.

\begin{figure}
\includegraphics[width=7.0cm]{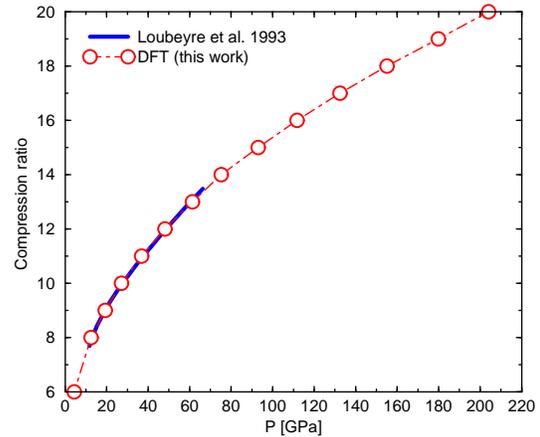}
\caption{Cold curve for solid helium in the h.c.p. phase. 
The static compression ratio, $V_0/V_1$ with $V_0=32.4$ cm$^3$/mol, is
plotted as function of pressure.  The experimental results~\cite{Lo93}
were obtained with diamond anvil cell measurement and compare
favorably with static density functional calculations performed at T=0
under hydrostatic conditions. }
\label{cold_curve} 
\end{figure}

The static DFT calculations used to derive the cold curve in
Fig.~\ref{cold_curve} were performed with 6x6x6 k-point mesh in a two
atom h.c.p. unit cell under hydrostatic conditions.
\section{RESULTS}
Figure~\ref{jupiter} shows a family of shock Hugoniot curves up to a
20-fold precompression. We characterize the precompression in terms of
volume change compared to the ambient pressure value of $V_0$=32.4
cm$^3$/mol~\cite{Ne83}. While static diamond anvil cell experiments
have explored the pressures beyond 300 GPa~\cite{Lo02}, there are
currently a number of limitations for the precompression pressure in
shock experiments. Larger pressures imply thicker diamond windows that
require a more powerful shock driver and make it more challenging to
launch a planar shock. Reducing the sample size in order to reach a
higher precompression makes the diagnostics more difficult.

\begin{figure}
\includegraphics[width=7.0cm]{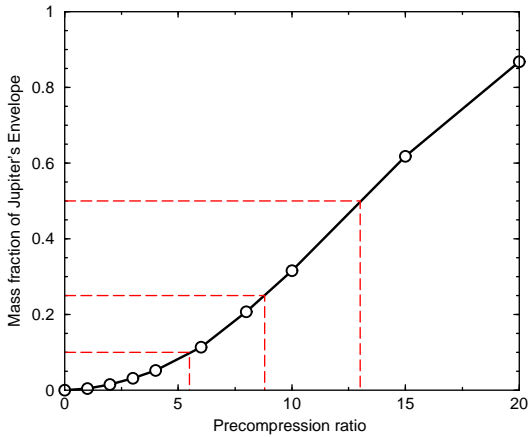}
\caption{Mass fraction of Jupiter's envelop that can be probed with different 
precompressions is shown as a function the precompression ratio, $\eta=V_0/V_1$. 
The dashed lines were included as guide to the eye.}
\label{jupiter_vs_ratio} 
\end{figure}

Let us now assume that we have a particular experimental setup that
allows us to launch shocks into a sample that has been precompressed
up a maximum initial pressure $P_1^*$. This translates into a maximum
precompression ratio, $\eta^*,$ that can be inferred from the cold
curve~\cite{Lo93,driessen86,Zha04} shown in Fig.~\ref{cold_curve} where we
compared our static DFT calculation with experimental results. In
Fig.~\ref{jupiter}, we find the pressure, $P_2^*$, where this
particular Hugoniot curve intersects with Jupiter's isentrope, and can
therefore infer the maximum depth in Jupiter's envelope that we can
probe with this particular experimental setup. Since it should always
be easier to repeat the experiment for smaller precompression, we can
map out Jupiter's isentrope for all $P_2<P_2^*$.

One can now ask the question how deep into Jupiter's interior one is
able to probe. However, Jupiter is an oblate object due to its rapid
rotation, and $P_2^*$ cannot be mapped directly into a radius without
approximating the planet by a sphere. It is therefore more appropriate
to ask what mass fraction of the total gas in the envelop is at
pressures less than $P_2^*$. This way one can map a maximum initial
pressure $P_1^*$ into a mass fraction of the planet that can be probed
experimentally. Discussing this in terms of the a mass fraction,
rather then in terms of depth, is particularly meaningful because
giant planet interior models are most sensitive to the equation of
state where most of the planet's mass is. For Jupiter, this is where
hydrogen is metallic and consequently a large number experimental and
theoretical studies have been devoted to hydrogen under such extreme
conditions.

\begin{figure}
\includegraphics[width=7.0cm]{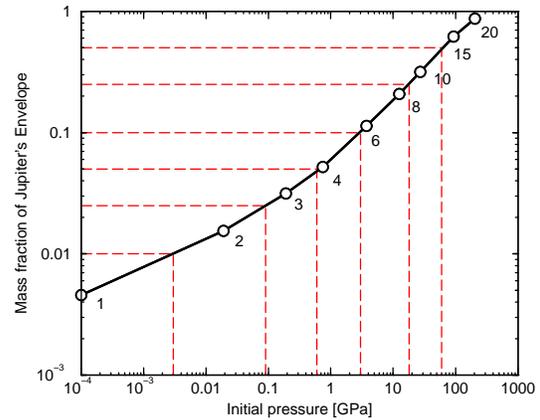}
\caption{Mass fraction of Jupiter's envelop that can be probed with different 
precompressions is shown as a function of the initial pressure. The dashed lines were 
included as guide to the eye. The labels give the precompression ratio, $V_0/V_1$, for each circle.}
\label{jupiter_vs_p} 
\end{figure}

Figures~\ref{jupiter_vs_ratio} and \ref{jupiter_vs_p} show the mass
fraction as function of precompression ratio and as function of
initial pressure, $P_1^*$. It becomes clear that a substantial
precompression pressure is needed to study Jupiter. With
precompressions of up to 0.1, 1.0, 10, or 100 GPa are needed to
characterized 2.5, 5.9, 18, to 63\% of Jupiter's envelop by
mass. While a precompression up 1 GPa increase the sample density
approximately 5-fold, it is not sufficient to characterize more than
5.9\% of Jupiter's mass. Since precompression above 1 GPa cannot
readily be obtained yet, one might consider performing double or
triple shock experiments~\cite{Kn04} with precompressed samples.

\section{CONCLUSIONS}

We performed density functional molecular dynamics simulations to
characterize the fluid helium at high pressure and temperature. We
derived the shock Hugoniot curve up to precompression ratio of 20. By
comparing the Hugoniot curve with Jupiter's isentrope, we conclude
that precompressions of up one megabar would be needed to characterize
a substantial fraction of Jupiter's envelop. Such large precompression
could be difficult to obtain due to limitations of shock drive and a
minimum sample size that is required. As a conclusion, we suggest that
venue of precompressed shock experiments with two or more
reverberating shocks should be explored.

\begin{theacknowledgments}
This material is based upon work
supported by NASA under the grant NNG05GH29G and by the NSF under the
grant 0507321.
\end{theacknowledgments}


\begin{thebibliography}{10}
\providecommand{\enquote}[1]{``#1''}
\expandafter\ifx\csname url\endcsname\relax
  \def\url#1{\texttt{#1}}\fi
\expandafter\ifx\csname urlprefix\endcsname\relax\def\urlprefix{URL }\fi

\bibitem{Si97}
Silva, L. B.~D., Celliers, P., Collins, G.~W., Budil, K.~S., Holmes, N.~C.,
  Jr., T. W.~B., Hammel, B.~A., Kilkenny, J.~D., Wallace, R.~J., Ross, M., and
  Cauble, R., \emph{Phys. Rev. Lett.}, \textbf{{78}}, 483 (1997).

\bibitem{Co98}
Collins, G.~W., Silva, L. B.~D., Celliers, P., Gold, D.~M., Foord, M.~E.,
  Wallace, R.~J., Ng, A., Weber, S.~V., Budil, K.~S., and Cauble, R.,
  \emph{Science}, \textbf{{281}}, 1178 (1998).

\bibitem{Kn01}
Knudson, M.~D., Hanson, D.~L., Bailey, J.~E., Hall, C.~A., Asay, J.~R., and
  Anderson, W.~W., \emph{Phys. Rev. Lett.}, \textbf{87}, 225501 (2001).

\bibitem{Kn03}
Knudson, M.~D., Hanson, D.~L., Bailey, J.~E., Hall, C.~A., and Asay, J.~R.,
  \emph{Phys. Rev. Lett.}, \textbf{90}, 035505 (2003).

\bibitem{Ze66}
Zeldovich, Y.~B., and Raizer, Y.~P., Academic Press, New York, 1966.

\bibitem{MC00}
Militzer, B., and Ceperley, D.~M., \emph{Phys. Rev. Lett.}, \textbf{85}, 1890
  (2000).

\bibitem{Mi06}
Militzer, B., \emph{Phys. Rev. Lett.}, \textbf{97}, 175501 (2006).

\bibitem{Lee06}
Lee, K. K.~M., Benedetti, R., Jeanloz, R., Celliers, P.~M., Eggert, J.~H.,
  Hicks, D.~G., Moon, S.~J., Mackinnon, A., DaSilva, L.~B., Bradley, D.~K.,
  Unites, W., Collins, G.~W., Henry, E., Koenig, M., Benuzzi-Mounaix, A.,
  Pasley, J., and Neely, D., \emph{J. Chem. Phys.}, \textbf{125}, 014701
  (2006).

\bibitem{VASP}
G. Kresse and J. Hafner, {\it Phys. Rev. B} 47, 558 (1993); G. Kresse and J.
  Hafner, {\it Phys. Rev. B} 49, 14251 (1994); G. Kresse and J. Furthm\"uller,
  {\it Comput. Mat. Sci.} 6, 15 (1996); G. Kresse and J. Furthm\"uller, {\it
  Phys. Rev. B} 54, 11169 (1996).

\bibitem{PBE}
Perdew, J.~P., Burke, K., and Ernzerhof, M., \emph{Phys. Rev. Lett.},
  \textbf{77}, 3865 (1996).

\bibitem{Vo07b}
Vorberger, J., Tamblyn, I., Bonev, S., and Militzer, B., \emph{Contrib. Plasma
  Phys.}, \textbf{47}, 375 (2007).

\bibitem{Vo07}
Vorberger, J., Tamblyn, I., Militzer, B., and Bonev, S., \emph{Phys. Rev. B},
  \textbf{75}, 024206 (2007).

\bibitem{Lo93}
Loubeyre, P., LeToullec, R., Pinceaux, J.~P., Mao, H.~K., Hu, J., and Hemley,
  R.~J., \emph{Phys. Rev. Lett.}, \textbf{71}, 2272 (1993).

\bibitem{Ne83}
Nellis, W., Mitchell, A., and {\it et al.}, \emph{J. Chem. Phys.}, \textbf{79},
  1480 (1983).

\bibitem{Lo02}
Loubeyre, P., Occelli, F., and LeToullec, R., \emph{Nature}, \textbf{416}, 613
  (2002).

\bibitem{driessen86}
Driessen, A., van~der Poll, E., and Silvera, I.~F., \emph{Phys. Rev. B},
  \textbf{33}, 3269 (1986).

\bibitem{Zha04}
Zha, C.~S., Mao, H.~K., and Hemley, R.~J., \emph{Phys. Rev. B}, \textbf{70},
  174107 (2004).

\bibitem{Kn04}
Knudson, M.~D., Hanson, D.~L., Bailey, J.~E., Hall, C.~A., Asay, J.~R., and
  Deeney, C., \emph{Phys. Rev. B}, \textbf{69}, 144209 (2004).

\end{thebibliography}
\end{document}